\documentstyle[11pt,newpasp,twoside,psfig,wrapfig]{article}
\index{Chandra}
\index{pulsar wind nebulae}
\index{pulsar wind nebulae!hydrodynamics}
\index{pulsar wind nebulae!time variability}
\index{bow shocks}
\index{shocks}
\index{shocks!reverse shock}
\index{supernova remnants!interactions with pulsar wind nebulae}
\index{pulsar wind nebulae!Crab Nebula}
\index{supernova remnants!B0453--685}
\index{pulsars!B1957+20}
\index{pulsar wind nebulae!Mouse}
\index{pulsars!B1509--58}

\def\edcomment#1{\iffalse\marginpar{\raggedright\sl#1\/}\else\relax\fi}
\reversemarginpar

\begin{document}
\title{Shocks and Wind Bubbles Around Energetic Pulsars}
\author{Bryan M.\ Gaensler}
\affil{Harvard-Smithsonian Center for Astrophysics, 60 Garden Street MS-6, Cambridge MA 02138, USA}

\begin{abstract}
The Crab Nebula demonstrates that 
neutron stars can interact with their environments in spectacular
fashion, their relativistic winds generating nebulae observable across
the electromagnetic spectrum.  At many previous conferences,
astronomers have discussed, debated and puzzled over the complicated
structures seen in the Crab, but have been limited to treating most
other pulsar wind nebulae (PWNe) as simple calorimeters for a pulsar's spin-down
energy.  However, with the wealth of high-quality data which have now become
available,
this situation has changed dramatically.  I here
review some of the main observational themes which have emerged from
these new measurements. Highlights
include the ubiquity of pulsar termination shocks,
the unambiguous presence of relativistic jets in PWNe,
complicated time variability seen in  PWN structures, and the use of
bow shocks to probe the interaction of pulsar winds with the
ambient medium.
\end{abstract}

\vspace{-6mm}
\section{Introduction}

Pulsars are thought to be born with initial spin periods in the range
10--100~ms, implying initial rotational kinetic energies of the order of
$\sim10^{50}$~ergs. Timing observations readily establish that all
isolated pulsars are spinning down; the implied {\em spin-down luminosity}\
is $\dot{E} = 4\pi^2I \dot{P}/{P^3}$, where $I\equiv10^{45}$~g~cm$^{-2}$
is the assumed moment of inertia for the neutron star, $P$ is the
star's rotational period and $\dot{P}$ is the spin period derivative.
For typical young pulsars, we find spin-down luminosities in the range
$\dot{E} \sim 10^{35}-10^{39}$~ergs~s$^{-1}$.

It is believed that most of this energy release ends up in a
relativistic particle wind (see Melatos, these proceedings). At the
interface where this wind interacts with its environment, a {\em
pulsar wind nebula}\ (PWN) is formed. PWNe form an ideal laboratory for
studying compact objects, relativistic shocks and particle acceleration,
all key themes in a wide range of important astrophysical problems.

\section{Overall Properties}

\begin{figure}
\centerline{
\psfig{file=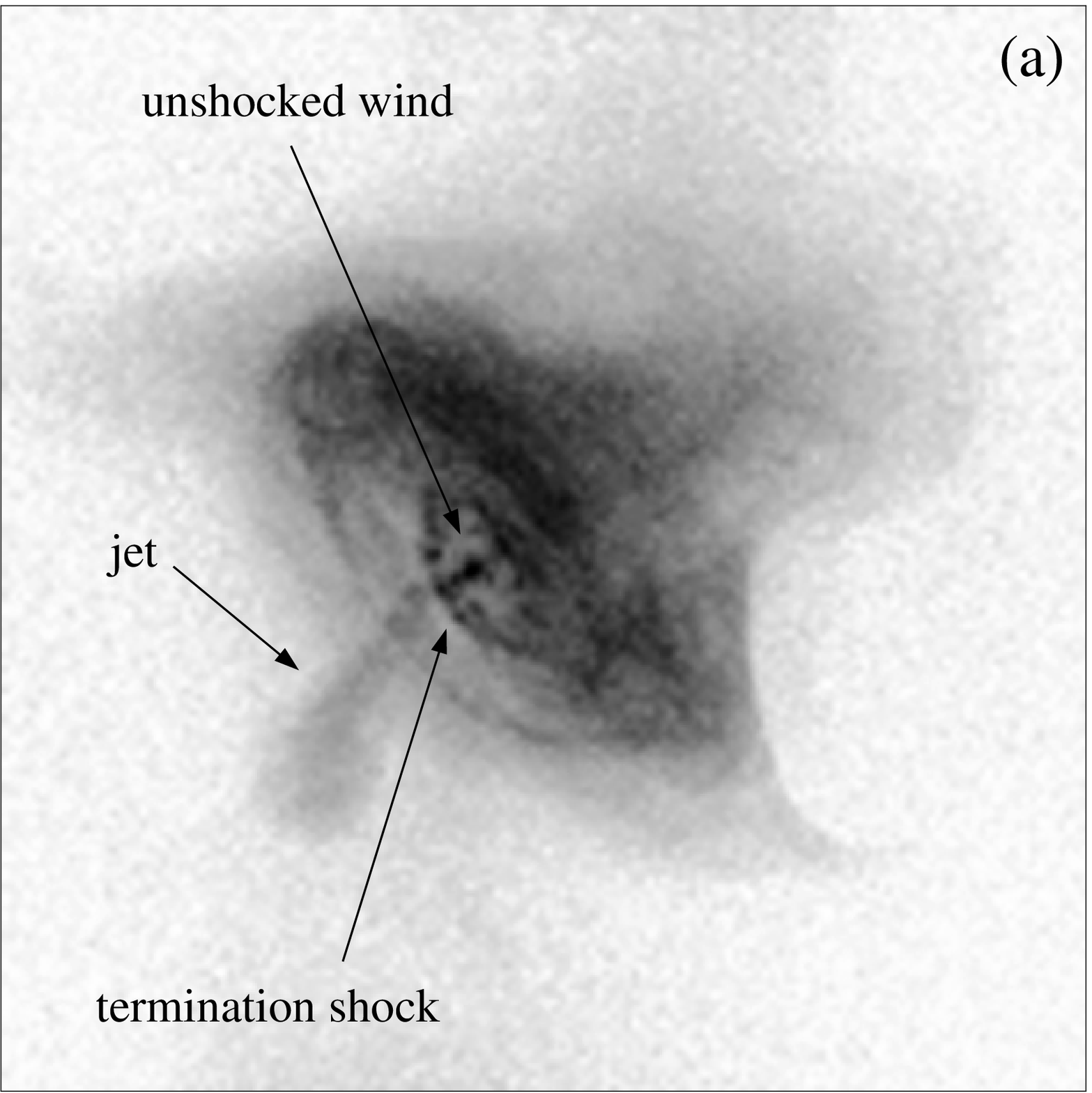,width=0.33\textwidth}
\psfig{file=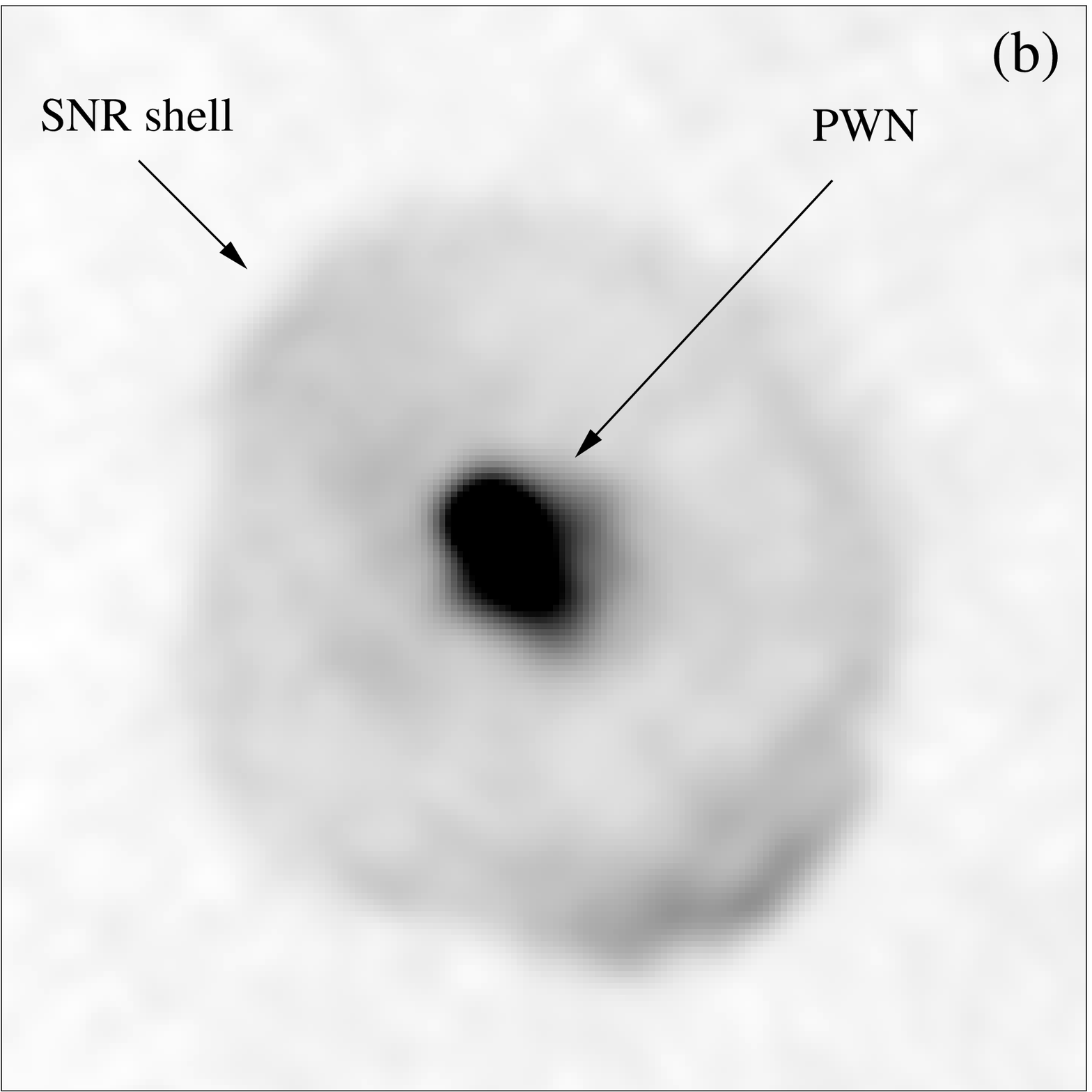,width=0.33\textwidth}
\psfig{file=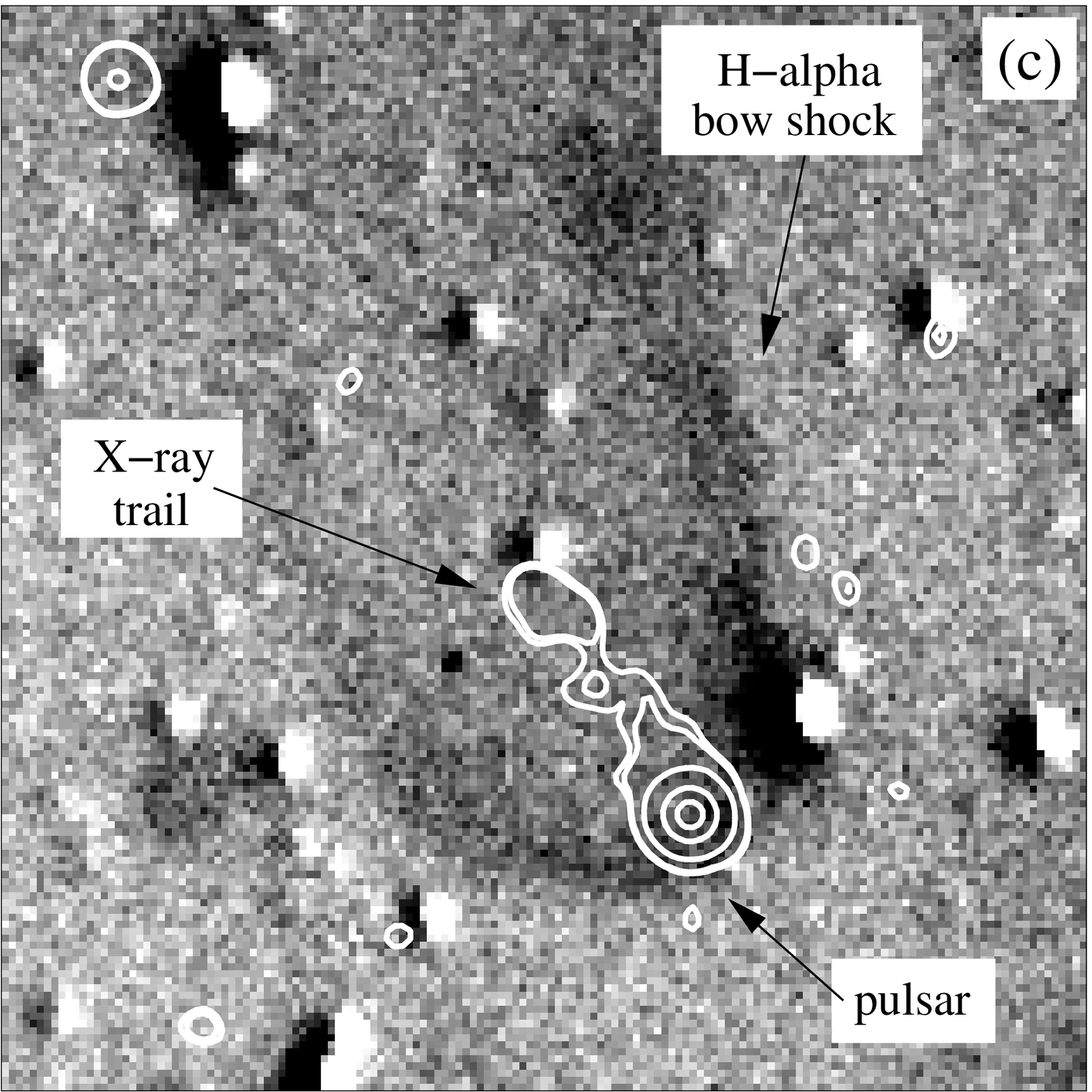,width=0.33\textwidth}}
\caption{Examples of the three broad phases in PWN evolution.
(a) {\em Chandra}\ X-ray image of the Crab Nebula;
(b) ATCA radio image of SNR~B0453--685 in the LMC;
(c) AAO H$\alpha$ (greyscale) and {\em Chandra}\ X-ray
(contours) images of the bow shock powered by PSR~B1957+20.}
\label{fig_pwne}
\vspace{-2mm}
\end{figure}

While there are wide variations in PWN properties, three broad
categories can be considered, loosely representing an evolutionary
sequence (Fig.~\ref{fig_pwne}; van der Swaluw \& Downes, these proceedings;
van der Swaluw et al.\ 2004)

In the youngest systems ($t \la 1000$~yr), $\dot{E}$ is approximately
constant as a function of time; the PWN expands supersonically into
unshocked low-density ejecta inside a surrounding expanding supernova
remnant (SNR).  In this phase the radius of the PWN is small but is
accelerating, with an approximate time-dependence $R \propto t^{6/5}$
(van der Swaluw et al.\ 2001, 2004).  PWNe likely to be in this phase
include the Crab Nebula (see Fig.~\ref{fig_pwne}[a]), 
and the PWN powered by PSR~J1811--1925 in the
young SNR~G11.2--0.3 (Roberts et al.\ 2003).

In middle-aged systems ($t \sim 10$--50~kyr), the interaction of the
expanding supernova ejecta with their surroundings produces both forward
and reverse shocks in the SNR. The latter collides with the pulsar
wind shock,
compressing and distorting the PWN. For
a stationary pulsar in a spherically symmetric system, the PWN now
expands subsonically. However, the effects of inhomogeneities in the
surrounding interstellar medium (ISM), and of the pulsar's 
space velocity, can produce a significantly distorted PWN,
with the pulsar near one end. Likely examples of PWNe in the process of
interacting with their SNR's reverse shock include G327.7--1.1 (van der Swaluw
\& Downes, these proceedings) and B0453--685 in the Large Magellanic Cloud
(Fig.~\ref{fig_pwne}[b]; Gaensler et al.\ 2003).

At later times ($t \ga 100$~kyr), the pulsar will move far from its
birth site, and will eventually escape its SNR. The pulsar's
motion is now supersonic, and its wind drives a bow shock, confined
by ram pressure.  Examples of bow shocks include those powered by
PSRs~B1957+20 (``the Black Widow'') (Fig.~\ref{fig_pwne}[c]; Stappers
et al.\ 2003) and J1747--2958 (``the Mouse'') (Fig.~\ref{fig_mouse}[b];
Gaensler et al.\ 2004).

\vspace{-4mm}
\section{Crab-like Pulsar Wind Nebulae}

In the earliest stages of evolution, a PWN is a quasi-spherical expanding
wind bubble with a constant central energy source.  Close to the pulsar,
we expect that wind particles flow freely outward in all directions. This
cold wind is not directly observable. At some distance (typically
$\sim0.1$~pc) from the pulsar, this wind is confined by external pressure,
and forms a {\em termination shock}.  Particles are accelerated at this
shock up to ultrarelativistic energies (see Arons, Lyubarsky,
these proceedings).
Downstream of the termination shock, the
flow further decelerates and the gyrating particles emit synchrotron
emission, forming the observable PWN.

For the Crab Nebula, the nebular magnetic field strength is high enough to
cause significant synchrotron losses at high energies. We thus observe a
centrally filled, linearly polarized source at all wavelengths, but whose
extent in the optical and X-ray bands is progressively smaller than seen
in the radio. It is sometimes claimed that at radio wavelengths (where
the synchrotron lifetimes are longer than the age of the PWN), the PWN's
luminosity traces the integrated history of the pulsar spin-down, while
in X-rays (where the radiative lifetimes are short), the PWN luminosity
traces the current energy output of the central star. However, while
this is broadly true for the Crab Nebula, there are other young PWNe
with much weaker magnetic fields, in which synchrotron losses are yet
to dominate in X-rays (e.g., the PWN around PSR~B1509--58; Gaensler
et al.\ 2002a). Furthermore, interpretation of a PWN's radio emission
as representing an integrated history of the system is complicated
by adiabatic losses and the effects of the reverse shock interaction
(Reynolds \& Chevalier 1984).

In cases where synchrotron losses are significant in X-rays, one expects
to see a spectral break somewhere between the radio and X-ray bands.
Indeed in some cases this is observed; the break frequency can then be
used to estimate the nebular magnetic field strength (e.g., Manchester
et al.\ 1993).  However, in other cases the field strength implied
by such a break is unphysically large (e.g., Green \& Scheuer 1992),
while other PWNe appear to have multiple breaks in their spectra (e.g.,
Bock \& Gaensler, these proceedings).  Clearly spectral
features in PWNe must be interpreted with caution, particularly when
one considers that breaks in the spectrum can also be introduced by
pulsar spin-down, or by intrinsic features in the injection spectrum
(e.g., Woltjer et al.\ 1997).

\subsection{Can We See Termination Shocks?}

As mentioned above, we expect that there is an inner unshocked wind zone
in young PWNe, bounded by a termination shock at which particles are
accelerated. Indeed, the superb angular resolution of {\em Chandra}\
has revealed in several PWNe an underluminous
region immediately surrounding the pulsar, surrounded by an X-ray
bright termination shock, as shown in Figure~\ref{fig_pwne}(a). This
clearly demarcates the point at which wind first interacts with its
surroundings. In many cases this shock has a ring-like, rather than a
spherical, geometry.  This demonstrates that the synchrotron-emitting
particles are concentrated into an equatorial flow around the pulsar
spin-axis (see discussion by Melatos, these proceedings).

Once we have identified this axial symmetry, the position angle and
eccentricity of this inner ring immediately provides the orientation
of the pulsar in three dimensions, which can prove extremely useful
in interpreting pulse profiles and in assessing the presence of
``spin-kick'' alignment (Ng \& Romani 2004).  Since the termination shock demarcates
the point where the wind and nebular pressures balance, we can write:
\begin{equation} 
\frac{\dot{E}}{\Omega r_w^2 c} = P_{nebula} \approx P_{fields} +
P_{particles}~, 
\label{eqn_bg_pressure} 
\end{equation}
where $\Omega$ is the solid angle of the wind, $r_w$ is the radius
of the termination shock in the equatorial plane and $P$ is the
pressure.  If we assume equipartition between fields and particles,
Equation~(\ref{eqn_bg_pressure}) then allows us to estimate the mean
nebular field strength. 
Brightness variations are observed around the rims of these inner rings, 
as expected
due to relativistic Doppler boosting. Fits to these intensity variations
allow us to directly determine the speed of the post-shock flow,
$v=\beta c$; typical
values are $0.4\la \beta \la 0.6$ (e.g.,
Ng \& Romani 2004)

\vspace{-1mm}
\subsection{Are Crab-like Pulsar Wind Nebulae Spherical?}

\begin{wrapfigure}{r}{6.5cm}
\vspace{-3mm}
\centerline{\psfig{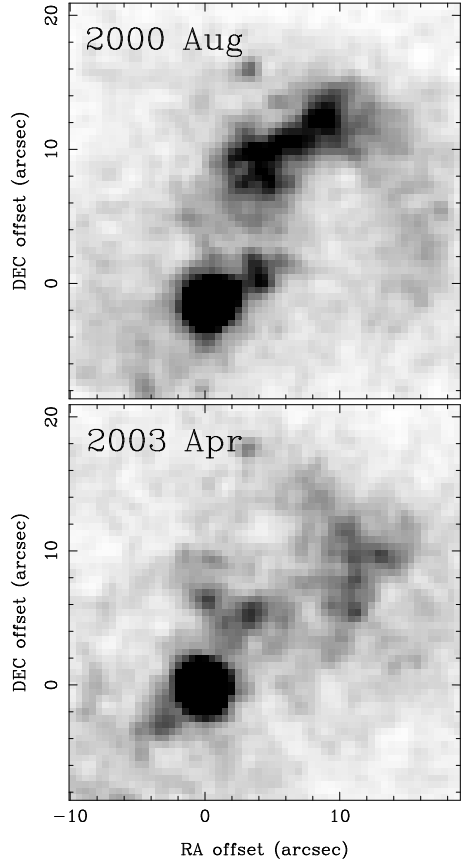}}
\caption{Multi-epoch {\em Chandra}\ images of the inner
regions of the PWN powered by PSR~B1509--58; the pulsar
is the bright source at the origin (Gaensler et al., in
preparation).}
\label{fig_b1509}
\end{wrapfigure}

Observations quickly demonstrate that young PWNe are not simple spherical
bubbles (e.g., Fig.~\ref{fig_pwne}[a]).  Many such sources show a clear
axial symmetry, with significant elongation around the centrally-located
pulsar. In cases where the termination shock can also be identified, this
elongation is always perpendicular to the inner equatorial ring. It thus
appears that the overall elongation of PWNe is an imprint of the pulsar
spin-axis at much larger scales; if so, PWN morphologies then provide
a simple way to infer the projected orientation of pulsar spin axes.
The overall elongation of young PWNe may result from the winding up
of nebular magnetic fields due to the pulsar's rotation, producing a
toroidal field geometry which affects the dynamics of the PWN's expansion
(Begelman \& Li 1992; van der Swaluw 2003). However, in some cases there
are striking collimated features, extending up to $\sim5$~pc from the
pulsar  (e.g., Gaensler et al.\ 2002a). Morphologically,
these features appear to be high-velocity jets of particles, directed
along the spin-axis. Indeed both proper motion  and spectral measurements
of these structures show them to have velocities $\beta \sim 0.3-0.6$
(Gaensler et al.\ 2002a; Hester et al.\ 2002).  It is not yet clear
what focuses these jets (see Melatos, these proceedings, for a detailed
discussion); Komissarov \& Lyubarsky (2003) have recently carried
out simulations which show that hoop stress in the downstream flow above
the spin axis may produce this collimation.

\vspace{-2mm}
\section{Time Variability}

Multi-epoch data in the radio, optical, infrared, X-ray and possibly
even gamma-ray bands all now demonstrate that PWNe show significant
time variability (e.g., Hester et al.\ 2002; Pavlov et al.\ 2003; Ling
\& Wheaton 2003; Melatos, these proceedings). Much of the most rapid
variability is observed near or even inside the termination shock 
(Fig.~\ref{fig_b1509}; Hester 1998), 
but complicated changes on timescales of months to years
are seen in the outer parts of PWNe also (Pavlov et al.\ 2003; Mori et
al., these proceedings).  We are only just beginning to characterize
these changes; clearly they represent a combination of phenomena,
corresponding to particle acceleration, streaming instabilities, standing
waves and turbulence. New theoretical
attempts to model the time-dependence of pulsar winds and PWNe should
provide important physical insight into the complicated processes being
traced by these data-sets.

\vspace{-2mm}
\section{Bow-Shock Pulsar Wind Nebulae}

Pulsar bow shocks provide a laboratory for studying pulsar winds
under a particularly well-defined geometry.  Furthermore, since pulsar
positions, distances, spin-down luminosities and velocities are often
well-constrained, these systems provide a physical situation which is
highly amenable to detailed analysis. Until recently, few bow-shock
systems were known. However, recent efforts have identified
a series of new systems (e.g., Olbert et al.\ 2001; Gaensler et al.\ 2002b).

\subsection{Theoretical Expectations}

\begin{figure}
\centerline{
\psfig{file=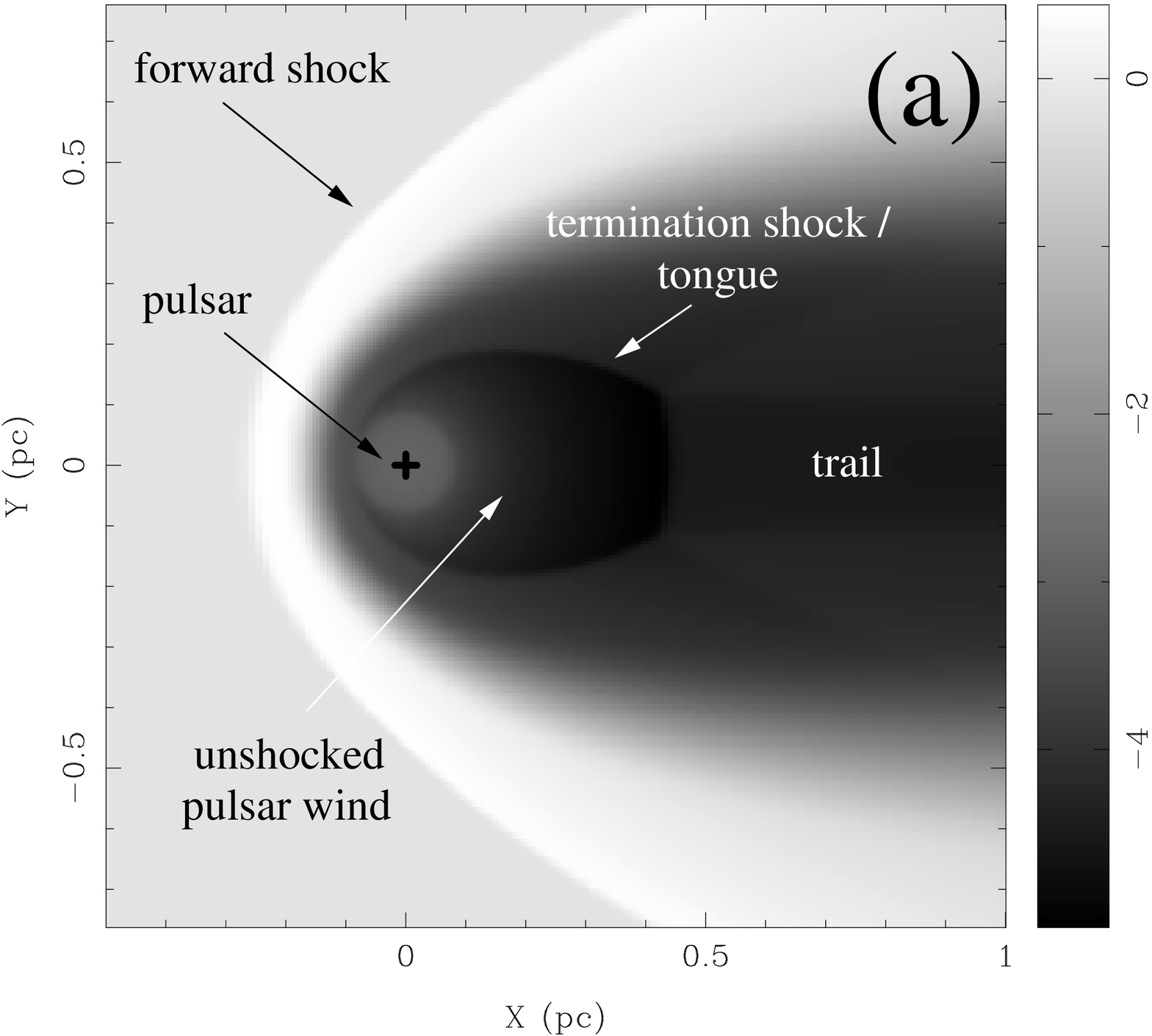,height=4.7cm,clip=}
\hspace{3mm}
\psfig{file=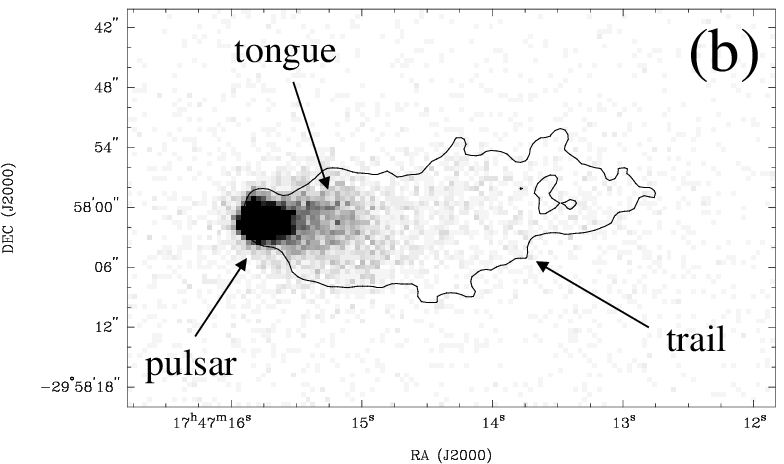,height=4.7cm,clip=}}
\caption{(a) Hydrodynamic simulation of a pulsar bow shock;
the greyscale represents density on a logarithmic scale.
(b) {\em Chandra}\ X-ray (greycale) and VLA radio (contour)
data on the bow shock powered by PSR~J1747--2958 (``the Mouse'').}
\label{fig_mouse}
\vspace{-4mm}
\end{figure}

An idealized bow shock can be simply understood. The ``stand-off
distance'' between the pulsar and the bow-shock apex is set by balance
between ram pressure and wind pressure:
\begin{equation}
\frac{\dot{E}}{\Omega r_w^2 c} = P_{ram} = \rho V^2,
\label{eqn_bow_balance}
\end{equation}
where $\rho$ is the mass density of the ambient medium and $V$ is the
pulsar's space velocity. Furthermore, once one has measured $r_w$,
the shape of the bow shock can be completely described by an analytic
solution:
\begin{equation}
r(\theta) = r_w / \sin \theta \sqrt{3(1-\theta/\tan \theta)},
\label{eqn_bow}
\end{equation}
where $r(\theta)$ is the distance of the bow shock from the pulsar at an
angle $\theta$ from the apex (Wilkin 1996).  In practice, a bow-shock PWN has a
double-shock structure, consisting of an outer forward shock and an inner
termination shock, separated by a contact discontinuity (see
Figure~\ref{fig_mouse}[a]). Simulations
suggest that the shape of the forward shock is still well-approximated
by Equation~(\ref{eqn_bow}).

\subsection{Optical Emission from Bow Shocks}

At the forward shock, we expect to see H$\alpha$ emission,
resulting from collisional excitation of neutrals in the ISM (e.g.,
Fig.~\ref{fig_pwne}[c]). Six bow shocks have been identified ---
five around radio pulsars and one around the isolated neutron star
RX~J1856.5--3754 (Chatterjee \& Cordes 2002; Gaensler et al.\ 2002b). If
$V$ is known, the measured stand-off distance can be used to infer
the ambient density via Equation~(\ref{eqn_bow_balance}).  However,
uncertainties in the inclination angle of the system ultimately limit
the accuracy of such estimates (Gaensler et al.\ 2002b).

For the bow shocks around PSR~J0437--4715 and RX~J1856.5--3754, the
morphology of the H$\alpha$ emission is a good match to the analytic
solution of Equation~(\ref{eqn_bow}) (e.g., van Kerkwijk
\& Kulkarni 2001). However, in the recently identified cases of
PSRs~B0740--28 and J2124--3358, there are significant deviations from
this geometry, implying some combination of anisotropies in the pulsar
wind flow (possibly corresponding to the torus/jet structures directly
imaged in younger systems; see Fig.~\ref{fig_pwne}[a])  and structure
in the ambient ISM (e.g., Gaensler et al.\ 2002b).

A typical pulsar space velocity of 500~km~s$^{-1}$ at a distance of 1~kpc
corresponds to a proper motion of $\sim0\farcs1$~yr$^{-1}$. The motion
of the corresponding bow shock is easily detectable in just a few years.
Indeed, Chatterjee \& Cordes (2004) have recently measured motion of the
``Guitar Nebula'' powered by PSR~B2224+65 over a seven-year baseline
with {\em HST}, revealing significant changes in the structure and
brightness of the optical emission from this PWN over this period. This
and other multi-epoch studies of optical bow shocks will provide a
unique measurement of fluctuations in the density of neutral gas in
the ISM on scales $\sim500-5000$~AU, filling a crucial gap between the
turbulent power spectrum seen in H\,{\sc i}\ on
scales $\sim0.1-200$~pc and the ``tiny scale atomic structure'' seen at
scales $\sim5-100$~AU.

\vspace{-2mm}
\subsection{Synchrotron Emission from Bow Shocks}

Just as for Crab-like PWNe, for
bow shocks we expect that the wind flows
freely close to the pulsar, before decelerating at a termination shock,
located inside the outer shock visible in H$\alpha$. Although the geometry
may differ from the Crab-like case, it is reasonable to expect that the
termination shock should also be a source of particle acceleration in
bow shocks, and that we should thus see a central region in X-ray and
radio synchrotron emission.

Indeed this has been confirmed in several sources, most notably around
PSRs~B1757--24 (``the Duck'') (Kaspi et al.\ 2001), B1853+01 (Petre et
al.\ 2002) and J1747--2958 (Fig.~\ref{fig_mouse}[b]; Gaensler et al.\
2004). Until recently there were no cases in which both the forward
and termination shocks had been identified in the same object.
Since only the most energetic pulsars produce bright radio and X-ray PWNe
(Gaensler et al.\ 2000; Gotthelf, these proceedings), and the radiation
from such pulsars might also ionize their surroundings, this is not overly
surprising (Chatterjee \& Cordes 2002).  However, in one case, that of the
recycled binary pulsar B1957+20, an H$\alpha$ bow shock and an enclosed
X-ray PWN are both now clearly seen (Fig.~\ref{fig_pwne}[c]; Stappers et
al.\ 2003). More instances of such systems need to be accumulated before
we can hope to understand what controls the observability of each shock
in such systems.

Observationally, the X-ray and radio emission from pulsar bow shocks
forms an elongated trail behind the pulsar (Figs.~\ref{fig_pwne}[c]
\& \ref{fig_mouse}[b]).  Only for PSR~B1957+20 have proper motion
measurements confirmed that this trail aligns with the pulsar's direction
of motion, but it is presumed that in other cases there is a similar
alignment.
Some authors have assumed these features to be synchrotron ``wakes''
left behind by the pulsar; in this case the length of the trail, combined
with the pulsar's velocity, puts a lower limit on the system's age.
However, in X-rays the radiative lifetimes are
far too short to account for the observed extent of these trails; e.g.\
for PSR~B1757--24, Kaspi et al.\ (2001) estimate a nebular magnetic field
strength of $\sim70$~$\mu$G, which should result in a trail $\la1''$ in
extent, in contrast to the $\sim20''$ trail observed. One possibility is
that there is a rapid flow behind the pulsar, which quickly transports
particles downstream to form the trail (Wang et al.\ 1993; Kaspi et
al.\ 2001).

A recent detailed comparison of {\em Chandra}\ and VLA data on the
Mouse suggests an alternative interpretation. The Mouse shows {\em two}\
components to its bright synchrotron trail: a bright narrow ``tongue''
seen close to the pulsar, and a fainter trail, seen at larger distances
downstream, as shown in Figure~\ref{fig_mouse}(b).  Gaensler et al.\
(2004) combine the theory of ion-dominated pulsar winds with hydrodynamic
simulations to show that the tongue likely represents the outer surface
of the wind termination shock, analogous to the inner ring seen in {\em
Chandra}\ images of the Crab Nebula, but stretched out due to ram pressure
from the pulsar's motion. The fainter elongated trail then corresponds
to synchrotron-emitting material seen further downstream.

It is important to note that 
the tongue feature seen for the Mouse resembles the {\em entire} trail
seen for systems such as PSRs~B1757--24 and B1957+20. I thus propose that
the trails seen in these bow shocks represent the bright termination
shock surrounding the pulsar, with the bulk of the PWN being much
larger and much fainter, presumably due to a combination of adiabatic
and synchrotron losses.  If this interpretation is correct, then there
is no need to invoke a rapid acceleration or collimation of the flow
behind the pulsar; all positions in the tongue represent locations of fresh
particle acceleration  in the relativistic flow.  Interestingly, {\em
Chandra}\ images of bow shocks around PSR~B1853+01 (Petre et al.\ 2002)
and CXOU~J061705.3+222127 (Olbert et al.\ 2001) both show suggestions of
a bright tongue surrounded by a fainter trail, as seen for the Mouse.
Deeper observations are needed to clarify these morphologies.

\section{Conclusions}

Some beautiful data-sets on PWNe are now letting us
address a number of fundamental questions regarding
these systems. We clearly now see the termination shock and post-shock
flow for Crab-like PWNe, and the double-shock structure expected for
bow-shock systems. We have learned how to use the morphology of these
systems to infer a pulsar's spin axis, 3D orientation, and space
velocity vector. We are beginning to explore the time-domain in PWN
studies, with which we can probe the dynamics of a pulsar's interaction
with its surroundings. While many questions remain to be answered,
we observers no longer need to apologize for our data,
and can happily
provide our theorist colleagues with
the measurements needed to properly address these issues.
Clearly a new era in the study of pulsars and their winds is now
well underway.

\acknowledgments I thank all my PWN collaborators for their enthusiastic
contributions to the work I have presented. My research on PWNe is
supported by NASA through SAO grants GO2-3079X, GO2-3074X, GO3-4068X,
GO2-3075X and GO3-4063A, XMM-Newton Guest Observer grants NAG5-11376,
NAG5-13087 and NAG5-13203,  and LTSA grant NAG5-13032.

\end{document}